\documentstyle[preprint,aps]{revtex}
\preprint{USM-TH-141}
\begin{document}
\title{The Cornell confining potential from spontaneous breaking of
scale symmetry}
\author{P. Gaete$^{1}$\thanks{
E-mail: patricio.gaete@fis.utfsm.cl} and E. I. Guendelman
$^{2}$\thanks{ E-mail: guendel@bgumail.bgu.ac.il}}
\address{$^1$Departamento de F\'{\i}sica, Universidad T\'{e}cnica
F. Santa Mar\'{\i}a, Casilla 110-V, Valpara\'{\i}so, Chile. \\
$^2$Physics Department, Ben Gurion University, Beer Sheva 84105,
Israel.} \maketitle
\begin{abstract}
We show that one can obtain naturally the Cornell confining
potential from the spontaneous symmetry breaking of scale
invariance in gauge theory. At the classical level a confining
force is obtained and at the quantum level, using a gauge
invariant but path-dependent variables formalism, the Cornell
confining potential is explicitly obtained.
\end{abstract}
\smallskip

PACS number(s): 11.10.Ef, 11.15.Kc

\section{Introduction}

The question of confinement in gauge theories has been approached
with the use of many different techniques and ideas, like lattice
gauge theory techniques\cite{Wilson} and non-perturbative
solutions of Schwinger-Dyson's equations\cite{Zachariesen}. All
these approaches have the goal of proving the existence of a
linear potential between static quark sources.

The study of the spectrum of heavy quark-antiquark systems is very
well understood. However, as is well known, the binding nergy of
an infinitely heavy quark-antiquark pair represents a fundamental
concept in $QCD$ which is expected to play an important role in
the understanding of quark confinement. In this respect we recall
that the famous "Cornell potential"\cite{Eichten} was postulated
in order to simulate the features of $QCD$, that is,
\begin{equation}
V =  - \frac{{\kappa}}{r} + \frac{r}{{a^2 }}, \label{Cornell}
\end{equation}
here $a$ is a constant with the dimensions of length.

It is worthwhile remarking at this point that the appearance of
the scale $a$ in the Cornell potential (\ref{Cornell}) is very
important. One should take notice that the original gauge field
theory does not have any scales. Furthermore gauge theories with
no scale have a symmetry which is associated to this, scale
invariance. Thus it follows that the confinement phenomena breaks
the scale invariance as the Cornell potential (\ref{Cornell})
explicitly shows by introducing the scale $a$.

In this paper we will investigate the connection between scale
symmetry breaking and confinement. In particular we will show the
appearance of the Cornell potential (\ref{Cornell}) after
spontaneous breaking of scale invariance in a specific
model\cite{Guendelman1}. The quark-antiquark potential is then
calculated using the gauge invariant variables
formalism\cite{Gaete1}.

We also draw attention to the fact that the scale invariant model
studied\cite{Guendelman1} introduces, in addition to the standard
gauge fields also maximal rank gauge field strengths of four
indices in four dimensions, $F_{\mu \nu \alpha \beta }  = \partial
_{\left[ \mu  \right.} A_{\left. {\nu \alpha \beta } \right]}$
where $ A_{\nu\alpha\beta} $ is a three index potential. The
integration of the equations of motion of the $ A_{\nu\alpha\beta}
$ field introduces a constant of integration $M$ which breaks the
scale invariance. As we will see, the linear term in the Cornell
potential arises from the constant of integration $M$. When $M=0$
the equations of motion reduce to those of the standard gauge
field theory.

A short note on the history of these kind of models and this way
of breaking scale invariance is in order here. This technique for
breaking scale invariance was used first in generally covariant
theories containing a dilaton field in
Refs.\cite{Guendelman2,Guendelman3}, in the context of a general
type of models which were studied (in non scale invariant form)
before\cite{Guendelman4}. This approach has also been used to
dynamically generate the tension of strings and
branes\cite{Guendelman5}. In
Refs.\cite{Guendelman2,Guendelman3,Guendelman4,Guendelman5} the
maximal rank gauge field strength derives from a potential which
is composite out of $D$-scalars.

In order to calculate the potential energy between a
quark-antiquark pair we will use the gauge-invariant but
path-dependent variables formalism\cite{Gaete1}. Here the
quark-antiquark state is made gauge invariant by the introduction
of a gauge field cloud which is basically the path-ordered
exponential of the gauge field potential along the path where the
two charges are located. This methodology has been used previously
in many examples for studying features of screening and
confinement in gauge theories \cite{Gaete2,Gaete3,Gaete4}.

\section{Scale invariance breaking and generation of confinement}

We will study the scale symmetry breaking in the context of an
Abelian theory. The non-Abelian generalization presents no
problems\cite{Gaete1}.

Our starting point is the well known action
\begin{equation}
S = \int {d^4 } x\left( { - \frac{1}{4}F_{\mu \nu } F^{\mu \nu } }
\right), \label{sib1}
\end{equation}
where $F_{\mu \nu }  = \partial _\mu  A_\nu   - \partial _\nu
A_\mu$. This theory is invariant under the scale symmetry
\begin{equation}
A_\mu  \left( x \right) \mapsto A_\mu ^ {\prime}  \left( x \right)
= \lambda A_\mu  \left( {\lambda x} \right), \label{sib2}
\end{equation}
here $\lambda$ is a constant.

Let us now rewrite (\ref{sib1}) with the use of an auxiliary field
$\omega$
\begin{equation}
S = \int {d^4 } x\left( { - \frac{1}{4}\omega ^2  +
\frac{1}{2}\omega \sqrt { - F_{\mu \nu } F^{\mu \nu } } }
\right).\label{sib3}
\end{equation}
From the equation of the $\omega$ field we get
\begin{equation}
\omega  = \sqrt { - F_{\mu \nu } F^{\mu \nu } }, \label{sib4}
\end{equation}
and replacing (\ref{sib4}) back into (\ref{sib3}) we get then
(\ref{sib1}). Substituting (\ref{sib4}) in (\ref{sib3}) is a valid
operation because (\ref{sib4}) is a constraint equation. Under a
scale transformation $\omega$ transforms as
\begin{equation}
\omega  \mapsto \lambda ^2 \omega \left( {\lambda x}
\right).\label{sib5}
\end{equation}

Let us now introduce a charge in the theory (\ref{sib3}): we will
now keep the form (\ref{sib3}) but now $\omega$ will not be an
elementary field, rather $\omega$ will be given by
\begin{equation}
\omega  = \varepsilon ^{\mu \nu \alpha \beta } \partial _{\left[
\mu  \right.} A_{\left. {\nu \alpha \beta } \right]}. \label{sib6}
\end{equation}
Notice that we have introduced a new degree of freedom, the three
index potential and it generates the 4-index field strength
$F_{\mu \nu \alpha \beta }  \equiv \partial _{\left[ \mu  \right.}
A_{\left. {\nu \alpha \beta } \right]}$ , a  "maximal rank"  ( of
4-indices in 4-dimensions) field strength. In that case the
equation of motion of $A_{\nu\alpha\beta}$ is
\begin{equation}
\varepsilon ^{\gamma \delta \alpha \beta } \partial _\beta  \left(
{\omega  - \sqrt { - F^{\mu \nu } F_{\mu \nu } } } \right) = 0,
\label{sib7}
\end{equation}
which is integrated to give
\begin{equation}
\omega  = \sqrt { - F_{\mu \nu } F^{\mu \nu } }  + M .\label{sib8}
\end{equation}
The integration constant $M$ spontaneously breaks the scale
invariance, since both $\omega$ and $\sqrt { - F^{\mu \nu } F_{\mu
\nu } }$ transform as in Eq.(\ref{sib5}) but $M$ does not
transform. Notice that $M$ has the same dimensions as the field
strength $F_{\mu\nu}$, that is, dimensions of $\left( {length}
\right)^{ - 2}$. We further observe that the variation of the
$A_\mu$ field produces the following equation
\begin{equation}
\frac{\partial }{{\partial x^\mu  }}\left( {\omega \frac{{F^{\mu
\nu } }}{{\sqrt { - F_{\alpha \beta } F^{\alpha \beta } } }}}
\right) = \frac{\partial }{{\partial x^\mu  }}\left[ {\left(
{\sqrt { - F_{\alpha \beta } F^{\alpha \beta } }  + M}
\right)\frac{{F^{\mu \nu } }}{{\sqrt { - F_{\alpha \beta }
F^{\alpha \beta } } }}} \right] = 0, \label{sib9}
\end{equation}
as we will see in the next section, the introduction of the
unusual $M$ term leads to the generation of confinement. One may
suspect this because the consideration of the $M$ term alone is
known to lead to such behavior. In that case the equations of
motion are obtained from an action of the form
\begin{equation}
S = k\int {d^4 x\sqrt { - F_{\mu \nu } F^{\mu \nu } } } ,
\label{sib10}
\end{equation}
where $k$ is a constant. Such model leads to confinement, as shown
in Refs. \cite{Aurilia,Amer}, and to string solutions. Among other
properties it is known that electric monopoles do not exist
\cite{Amer}. We will see however that the consideration of the two
terms in (\ref{sib9}) leads to a richer structure in particular to
solutions containing Coulomb and linear parts, as in the Cornell
potential.

\section{Classical solutions and effective actions}

In order to illustrate the discussion, we now study the equation
(\ref{sib9}) for the case of a spherically symmetric electric
field $F_{0i}=-E_i$ and $F_{ij}=0$, where ${\bf E}=E(r)\hat {\bf
r}$. Then (\ref{sib9}) gives
\begin{equation}
\nabla  \cdot \left( {{\bf E} + \frac{M}{{\sqrt 2 }}\hat {\bf r}}
\right) = 0, \label{sib11}
\end{equation}
which is solved by
\begin{equation}
{\bf E} =  - \frac{M}{{\sqrt 2 }}\hat {\bf r} + \frac{q}{{r^2
}}\hat{\bf r}. \label{sib12}
\end{equation}
The scalar potential $V$ that gives rise to such electric field is
\begin{equation}
V  =  - \frac{M}{{\sqrt 2 }}r + \frac{q}{r}, \label{sib13}
\end{equation}
which is indeed resembles very much the Cornell potential
(\ref{Cornell}). Notice that so far (\ref{sib13}) refers to the
field of one charge and not yet to the interaction energy between
two charges. We will see that such interaction energy also has the
Cornell form, even at the quantum level. Since Abelian solutions
are solutions of the non-Abelian theory, these solutions are also
relevant for the non-Abelian generalization.

Before approaching the quantum theory (which will be treated in
some approximations) we want to define effective actions that give
the equations of motion (\ref{sib9}). Indeed one can easily see
that
\begin{equation}
{\cal L}_{eff}  =  - \frac{1}{4}F_{\mu \nu } F^{\mu \nu }  -
\frac{M}{4}\sqrt { - F_{\mu \nu } F^{\mu \nu } }, \label{sib14}
\end{equation}
reproduces Eqs. (\ref{sib9}).

Since the full treatment of the quantum theory is rather
difficult, instead of using (\ref{sib14}) we restrict ourselves to
a "truncated" phase space model where we consider spherical
coordinates $(r,\theta,\varphi)$ in addition to time, but where we
set $F_{ij}=0=F_{0\varphi}=F_{0\theta}$ and consider only $(t,r)$
dependence of $F_{0r}$. Then instead of (\ref{sib14}), we consider
\begin{equation}
S = 4\pi \int {dr} r^2 {\cal L}_{eff}, \label{sib15}
\end{equation}
where
\begin{equation}
{\cal L}_{eff}  = \frac{1}{2}\left( {F_{0r} } \right)^2  -
\frac{{M\sqrt 2 }}{4}F_{0r}. \label{sib16}
\end{equation}
Similar kind of "reduced phase space" which take into account only
the spherical degrees of freedom have been used elsewhere in other
examples, see for example Ref.\cite{Benguria}.

\section{Interaction energy}

As already mentioned, our aim now is to calculate the interaction
energy between external probe sources in the model (\ref{sib15}).
To do this, we will compute the expectation value of the energy
operator $H$ in the physical state $\left| \Phi  \right\rangle$,
which we will denote by $\left\langle H \right\rangle _\Phi$. The
starting point is the two-dimensional space-time Lagrangian
(\ref{sib15}):
\begin{equation}
{\cal L} = 4\pi r^2 \left\{ { - \frac{1}{4}F_{\mu \nu } F^{\mu \nu
} - \frac{{M\sqrt 2 }}{8}\varepsilon _{\mu \nu } F^{\mu \nu } }
\right\} - A_0 J^0, \label{poten1}
\end{equation}
where $J^0$ is the external current. A notation remark, in
(\ref{poten1}), $\mu,\nu=0,1$, also, $ x^1  \equiv r \equiv x$ and
$\varepsilon^{01}=1$.

We now proceed to obtain the Hamiltonian. For this we restrict our
attention to the Hamiltonian framework of this theory. The
canonical momenta read $\Pi ^\mu   =  - 4\pi x^2 \left( {F^{0\mu }
+ \frac{{M\sqrt 2 }}{8}\varepsilon ^{0\mu } } \right)$, which
results in the usual primary constraint $\Pi^0=0$, and $\Pi ^i = -
4\pi x^2 \left( {F^{0i}  + \frac{{M\sqrt 2 }}{8}\varepsilon ^{0i}
} \right)$. The canonical Hamiltonian following from the above
Lagrangian is:
\begin{equation}
H_C  = \int {dx} \left( {\Pi _1 \partial ^1 A^0  - \frac{1}{{8\pi
x^2 }}\Pi _1 \Pi ^1  - \frac{{M\sqrt 2 }}{4}\varepsilon ^{01} \Pi
_1  + A_0 J^0 } \right). \label{poten2}
\end{equation}
The consistency condition ${\dot \Pi _0}=0$ leads to the secondary
constraint $\Gamma _1 \left( x \right) \equiv \partial _1 \Pi ^1 -
J^0=0$. It is straightforward to check that there are no further
constraints in the theory, and that the above constraints are
first class. The extended Hamiltonian that generates translations
in time then reads $H = H_C  + \int d x \left( {c_0 (x)\Pi_0 (x) +
c_1 (x)\Gamma _1 (x)} \right)$, where $c_0(x)$ and $c_1(x)$ are
the Lagrange multipliers. Moreover, it follows from this
Hamiltonian that $ \dot{A}_0 \left( x \right) = \left[ {A_0 \left(
x \right),H} \right] = c_0 \left( x \right)$, which is an
arbitrary function. Since $\Pi_0 = 0$, neither $A^0$ nor $\Pi^0$
are of interest in describing the system and may be discarded from
the theory. The Hamiltonian then takes the form
\begin{equation}
H = \int {dx} \left( { - \frac{1}{{8\pi x^2 }}\Pi _1 \Pi ^1  -
\frac{{M\sqrt 2 }}{4}\varepsilon ^{01} \Pi _1  + c^ \prime  \left(
{\partial _1 \Pi ^1  - J^0 } \right)} \right), \label{poten3}
\end{equation}
where $c^ \prime  \left( x \right) = c_1 \left( x \right) - A_0
\left( x \right)$.

According to the usual procedure we introduce a supplementary
condition on the vector potential such that the full set of
constraints becomes second class. A convenient choice is found to
be \cite{Gaete1,Gaete2,Gaete3,Gaete4}
\begin{equation}
\Gamma _2 \left( x \right) \equiv \int\limits_{C_{\xi x} } {dz^\nu
} A_\nu \left( z \right) \equiv \int\limits_0^1 {d\lambda x^1 }
A_1 \left( {\lambda x} \right) = 0, \label{poten4}
\end{equation}
where  $\lambda$ $(0\leq \lambda\leq1)$ is the parameter
describing the spacelike straight path $ x^1  = \xi ^1  + \lambda
\left( {x - \xi } \right)^1 $, and $ \xi $ is a fixed point
(reference point). There is no essential loss of generality if we
restrict our considerations to $ \xi ^1=0 $. In this case, the
only nontrivial Dirac bracket is
\begin{equation}
\left\{ {A_1 \left( x \right),\Pi ^1 \left( y \right)} \right\}^ *
= \delta ^{\left( 1 \right)} \left( {x - y} \right) -
\partial _1^x \int\limits_0^1 {d\lambda x^1 } \delta ^{\left( 1
\right)} \left( {\lambda x - y} \right). \label{poten5}
\end{equation}

We are now equipped to compute the interaction energy between
pointlike sources in the model (\ref{sib15}), where a fermion is
localized at the origin $ {\bf 0}$ and an antifermion at $ {\bf
y}$. As we have already mentioned, we will calculate the
expectation value of the energy operator $H$ in the physical state
$ |\Phi\rangle$. From our above discussion, we see that
$\left\langle H \right\rangle _\Phi$ reads
\begin{equation}
\left\langle H \right\rangle _\Phi   = \left\langle \Phi
\right|\int {dx} \left( { - \frac{1}{{8\pi x^2 }}\Pi _1 \Pi ^1  -
\frac{{M\sqrt 2 }}{4}\varepsilon ^{01} \Pi _1 } \right)\left| \Phi
\right\rangle . \label{poten6}
\end{equation}
Next, as remarked by Dirac\cite{Dirac}, the physical state can be
written as
\begin{equation}
\left| \Phi  \right\rangle  \equiv \left| {\overline \Psi  \left(
\bf y \right)\Psi \left( \bf 0 \right)} \right\rangle  = \overline
\psi \left( \bf y \right)\exp \left( {ie\int\limits_{\bf 0}^{\bf
y} {dz^i } A_i \left( z \right)} \right)\psi \left(\bf 0
\right)\left| 0 \right\rangle, \label{poten7}
\end{equation}
where $\left| 0 \right\rangle$ is the physical vacuum state. As we
have already indicated, the line integral appearing in the above
expression is along a spacelike path starting at $\bf 0$ and
ending $\bf y$, on a fixed time slice.

Taking into account the above Hamiltonian structure, we observe
that
\begin{equation}
\Pi _1 \left( x \right)\left| {\overline \Psi  \left( y
\right)\Psi \left( 0 \right)} \right\rangle  = \overline \Psi
\left( y \right)\Psi \left( 0 \right)\Pi _1 \left( x \right)\left|
0 \right\rangle  - e\int_0^y {dz_1 } \delta ^{\left( 1 \right)}
\left( {z_1  - x} \right)\left| \Phi  \right\rangle.
\label{poten8}
\end{equation}
Inserting this back into (\ref{poten6}), we get
\begin{equation}
\left\langle H \right\rangle _\Phi   = \left\langle H
\right\rangle _0  + \frac{{e^2 }}{{8\pi }}\int {dx} \frac{1}{{x^2
}}\left( {\int_0^y {dz_1 \delta ^{\left( 1 \right)} } \left( {z_1
- x} \right)} \right)^2  + \frac{{M\sqrt 2 e}}{4}\int {dx} \left(
{\int_0^y {dz_1 } \delta ^{\left( 1 \right)} \left( {z_1  - x}
\right)} \right), \label{poten9}
\end{equation}
where $\left\langle H \right\rangle _0  = \left\langle 0
\right|H\left| 0 \right\rangle$. We further note that
\begin{equation}
\frac{{e^2 }}{2}\int {dx} \left( {\int_0^y {dz\delta ^1 \left(
{z_1  - x} \right)} } \right)^2  = \frac{{e^2 }}{2}L ,
\label{pato}
\end{equation}
with $|y|\equiv L$. Inserting this into Eq.(\ref{poten9}), the
interaction energy in the presence of the static charges will be
given by
\begin{equation}
V =  - \frac{{e^2 }}{{8\pi }}\frac{1}{L} + \frac{{M\sqrt 2
e}}{4}L, \label{poten10}
\end{equation}
which has the Cornell form. In this way the static interaction
between fermions arises only because of the requirement that the
$\left| {\overline \Psi \Psi } \right\rangle$ states be gauge
invariant.

\section{Conclussions}

We have found that in the context of a model where scale
invariance is spontaneously broken, the Cornell confining
potential between quark-antiquark naturally appears. The solutions
appear also relevant to the non-Abelian generalizations of the
model. Once again, the gauge-invariant formalism has been very
economical in order to obtain the interaction energy, this time
showing a confining effect in $(3+1)$ dimensions. Other aspects of
$QCD$ concern gluon confinement, in addition to the
quark-antiquark confinement we have studied so far. Indeed,
preliminary studies indicate that Eq.(\ref{sib9}), do not support
plane wave solutions, which is a clear hint of gluon confinement.
We will report on these issues in a future publication. Finally,
it would also be interesting to see if this model can describe
other confined states, like baryons, glueballs, etc.

\section{ACKNOWLEDGMENTS}

One of us (E.I.G.) wants to thank the Physics Department of the
Universidad T\'{e}cnica F. Santa Mar\'{\i}a for hospitality. P.G.
would like to thank I. Schmidt for his support.

\end{document}